\documentclass[11pt]{article}
\usepackage{moriond,epsfig}
\usepackage{graphicx}

\def\Journal#1#2#3#4{{#1} {\bf #2}, #3 (#4)}
\def\apj{\em ApJ}
\def\apjl{\em ApJL}
\def\apjs{\em ApJS}
\def\aap{\em A\&A}
\def\aaps{\em A\&AS}
\def\ssr{\em SSR}
\def\mnras{\em MNRAS}
\def\pasp{\em PASP}

\def\na{\em New Ast.}
\def\rmaa{\em RMAA}
\def\rmaacs{\em RMAAC}
\def\pasj{\em PASJ}
\def\adndt{\em ADNDT}
\def\nat{\em Nature}

\def\npps{\em NPPS}
\def\phrd{\em PhRD}

\def\lesssim{\mathrel{\hbox{\rlap{\hbox{\lower4pt\hbox{$\sim$}}}\hbox{$<$}}}}
\def\greatsim{\mathrel{\hbox{\rlap{\hbox{\lower4pt\hbox{$\sim$}}}\hbox{$>$}}}}

\begin{document}
\vspace*{4cm}
\title{THE PRIMORDIAL HELIUM ABUNDANCE}

\author{V. LURIDIANA}

\address{Instituto de Astrof\'\i sica de Andaluc\'\i a (CSIC),
Camino Bajo de Hu\'etor 24, 18008 Granada, Spain}

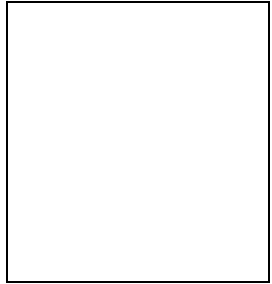
\maketitle\abstracts{
The primordial abundance of $^4$He, $Y_P$, is one of the hottest themes
in present-day astronomy, mostly due to its cosmological relevance.
The disagreement between different determinations has been currently reduced to 
the 1-2\% level, but these differences are still large enough 
to have deep implications for Big-Bang nucleosynthesis. 
It is therefore crucial to estimate precisely the uncertainties
involved in the measurement of $Y_P$.
Here, I review the methods used in the determination of $Y_P$
and the related uncertainties. 
I also discuss some recent results, and emphasize
the assumptions underlying the differences among them.
} 


\section{Introduction}
Research on the abundance of primordial helium ($Y_P$) 
has entered its golden age.
The exact value of $Y_P$
is one of the few missing pieces in the puzzle of the Big-Bang scenario, 
fueling a lively debate among astronomers,
and motivating a huge amount of literature on the subject.
The other side of the coin 
is that this literature 
is stuffed with numerically subtle argumentations,
devoted to the quantitative determination
of very specific aspects of the issue.
Having established the basic principles, all the effort is now committed
to the fine tuning of these details.
$Y_P$ measurements are increasingly precise, 
but still not accurate enough 
to be used as strong cosmological constraints.
The debate concerns now the third digit of $Y_P$:
a scale small indeed, but where differences still matter.

My scope here is to present an overview of the subject
and its implications. The most important part of this task
is trying to help the reader understand,
without getting lost in numbers, where the differences come from.
Unfortunately, 
I find it impossible to be simultaneously clear 
and exhaustive in just a few pages. My own choice is to be as clear as I can, though it implies that many important studies on this topic
will not be mentioned. I apologize for that to their authors,
hoping that this review will have, at least, the effect to make readers 
grasp the essence of the debate, and, hopefully, want to know more.

\section{Cosmological implications\label{sec:cosmo}}
\begin{figure}
  {\includegraphics{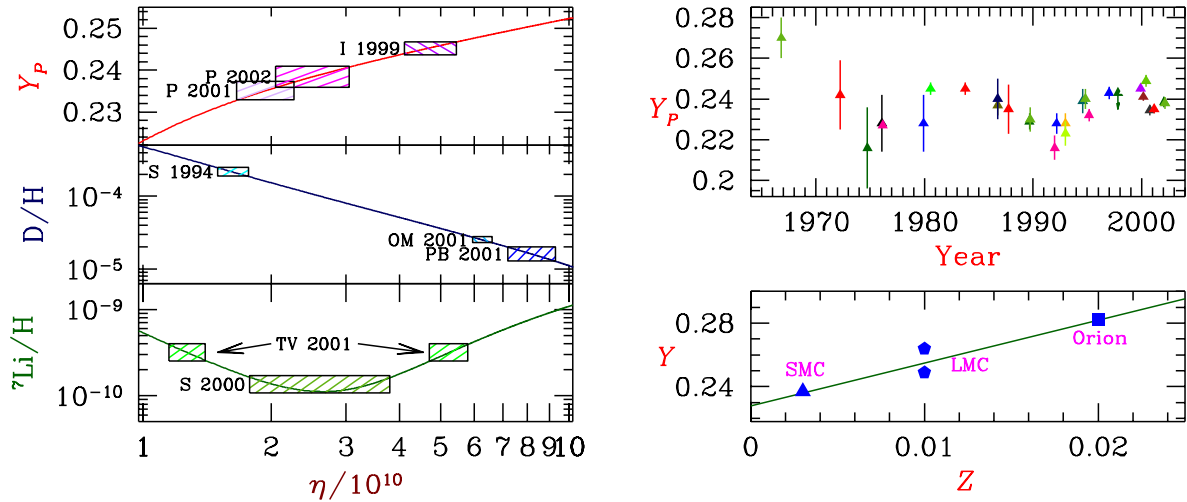}}
\caption{Left: baryon-to-photon ratio $\eta$ as a function of the primordial elemental abundances. The curves follow the relations by Fiorentini {et al.} ([11]), the abundance determinations are from [37], [38], [17], [59], [30], [48], [65], and [64]. Top right: $Y_P$ determinations plotted against publication year. The data points are from (left to right) [36], [54], [44], [45], [7], [24], [12], [23], [35], [22], [62], [66], [34], [26], [33], [25], [58], [19], [28], [20], [15], [18], [17], [68], [1], [42], [37], and [38]. Bottom right: the {\sl (Y, Z)} relation by Peimbert \& Torres-Peimbert ([45]).} 
\label{fig:yp}
\end{figure}

Primordial nucleosynthesis is one of the
three pillars supporting the Big-Bang model 
for the origin of the Universe,
the other two being the cosmic microwave background 
and the Hubble expansion
(Schramm~\cite{S98}; Dolgov~\cite{D02}). 
Since the Hubble expansion is also predicted by alternative
cosmological models, 
the Big-Bang nucleosynthesis (BBN) has a fundamental role 
as a decisive proof of the Big Bang.
In the scenario of standard BBN,
the primordial abundances of four light isotopes 
(D, $^3$He, $^4$He, and $^7$Li) 
depend only on the baryon-to-photon ratio $\eta$;
the corresponding relations for three of them are shown in Fig.~\ref{fig:yp}.
$\eta$ is a key parameter in cosmology, since it allows determination
of $\Omega_b$, the baryon fraction of the closure density. 
$\eta$ is overdetermined by the four isotopes,
therefore their agreement provides an extremely robust test of BBN,
and the four are actively investigated.
Each isotope tells a different history,
and represents a unique technical and theoretical challenge
(see, e.g., the review by S.~Burles in these proceedings).
Unfortunately, the data obtained in this field
are still contradictory and uncertain.
As for $^4$He, it
is the easiest to observe among the four,
but also the less sensitive to $\eta$,
so that its measurements must be highly precise
to be cosmologically relevant.
In the following, I will describe how these measurements are carried out,
highlighting the uncertainties involved in the analysis, 
and the causes of disagreement among different determinations.

\section{Y$_P$ determinations: the method}

The history of Y$_P$ determinations began in 1966 with Peebles\cite{P66},
who estimated {$0.26<Y_P<0.28$}, based on a simple cosmological model.
Many $Y_P$ determinations have followed since; some of them are
plotted in the upper right panel of Fig.~\ref{fig:yp}
as a function of the publication date.
The plot shows that the data are progressively converging toward a common value,
but also that a significant scatter remains, even among the most recent results.
In particular, the data published in the last five years 
tend to cluster around two preferred values 
($Y_P\sim 0.238$ and $Y_P\sim 0.245$), which
are, with their error bars, mutually exclusive.
Intermediate results are also found.

Before analyzing the causes underlying this disagreement,
I will give a general description of the method used to determine $Y_P$.
The basic strategy has been originally proposed by
Peimbert \& Torres-Peimbert~\cite{PTP74}:
since the Universe was born with zero metallicity ($Z$), and both $Z$
and the helium abundance ($Y$) increase with time, $Y_P$ can be found by extrapolating 
back to $Z$=0 the {($Y, Z$)} relation for a sample of objects
(Fig.~\ref{fig:yp}, lower right panel).
Variations of this basic strategy also exist:
e.g., $Y_P$ (or, more precisely, an upper limit to it)
can also be found by averaging $Y$ in a sample of extremely 
metal-poor objects (Searle \& Sargent~\cite{SS72}; Steigman, Viegas, \& Gruenwald~\cite{SVG97}).

The application of this method relies on precise measurements of $Y$ and $Z$
in individual objects.
This is done by means of nebular abundance diagnostics,
which are relations linking the observed emission line intensities
to the corresponding ionic abundances in the gas.
Many types of objects have been used in this analysis:
planetary nebulae (D'Odorico, Peimbert, \& Sabbadin~\cite{DPS76}),
H{\sc~ii} regions, either galactic  (Peimbert \& Torres-Peimbert~\cite{PTP74,PTP76}),
Magellanic 
(Peimbert \& Torres-Peimbert~\cite{PTP74,PTP76}; 
Peimbert, Peimbert, \& Luridiana~\cite{PPL02}),
or extragalactic (Skillman~\cite{S89}; 
Torres-Peimbert, Peimbert, \& Fierro~\cite{TPPF89}),
dwarf irregulars (dIrrs) and blue compact galaxies (BCGs) 
(Lequeux {\sl et al.}~\cite{Lal79}; Kunth~\cite{K86}; Pagel, Terlevich, \& Melnick~\cite{PTM86}; Skillman \& Kennicutt~\cite{SK93}; 
Izotov, Thuan, \& Lipovetsky~\cite{ITL94,ITL97}; Izotov \& Thuan~\cite{IT98,Ial99}; 
Fields \& Olive~\cite{FO98}; Peimbert {\sl et al.}~\cite{PPL02}).

Since $Z$ cannot, in practice, be directly measured,
individual metals are used as metallicity tracers;
the expected behavior of the element with respect to $Y$
determines the type of fit to the data.
Occasionally, carbon has been used for this scope
(see, e.g., Steigman~\cite{S87}, and Fields \& Olive~\cite{FO98})
but it does not yield, for various reasons, a good determination of $Y_P$.
Nitrogen is more frequently used; the relation between $Y$ and $N$
is sometimes assumed to be linear, 
but this assumption is an oversimplification,
roughly valid only at low $Z$
(see, e.g., Fields \& Olive~\cite{FO98}; van Zee, Salzer, \& Haynes~\cite{vZSH98}).
On the other hand, the linear behavior is a much better assumption
in the case of oxygen.
Oxygen, the most abundant heavy element, is the best metallicity tracer.

The slope in the ($Y$, $Z$) relation
can be determined either observationally
(i.e., from the fit to the data: Peimbert \& Torres-Peimbert~\cite{PTP74};
D'Odorico {\sl et al.}~\cite{DPS76}; 
Melnick {\sl et al.}~\cite{MHML92};
Pagel {\sl et al.}~\cite{Pal92a};
Izotov {\sl et al.}~\cite{ITL94,ITL97};
Izotov \& Thuan~\cite{IT98};
see also Pagel \& Portinari~\cite{PP98} and 
H{\o}g {\sl et al.}~\cite{Hal98} for a different approach),
or by means of chemical evolution models
(Lequeux {\sl et al.}~\cite{Lal79}; 
Carigi {\sl et al.}~\cite{Cal95}).
Most derived values are in the range $\sim 2-6$, with large uncertainties.
Obviously, the impact on $Y_P$ of the uncertainty in d$Y$/d$Z$
is minimized by the inclusion of BCGs and dIrrs in the sample, 
since these are the most metal-poor galaxies known.
On the other hand, d$Y$/$dZ$
is better determined with
a wide baseline in metallicity.

Helium abundance determinations are most often made by means
of optical observations,
although data from other wavelength ranges have also been used, 
e.g. infrared (Rubin {\sl et al.}~\cite{Ral98}) and radio 
(Churchwell, Mezger, \& Huchtmeier~\cite{CMH74}; Shaver {\sl et al.}~\cite{S83};
Peimbert {\sl et al.}~\cite{PUH88,Pal92b}).
Helium in photoionized regions 
can exist in all its three ionization stages.
Neutral helium cannot be observed, and will be dealt with in the next section.
Double-ionized helium, if present, gives rise to the
He II recombination spectrum, which is straightforward to interpret 
in terms of abundance.
None of these two ions is abundant in H{\sc ii} regions:
most of the helium is always singly ionized, 
and shows up in the spectrum as prominent He{\sc~i} recombination lines.
He{\sc~i} has two separate level systems, the singlets 
and the triplets, and the transitions 
between them are forbidden by electric dipole selection rules.
While singlet lines are relatively easy to interpret,
the triplet spectrum is complicated by the metastability
of the lowest triplet level, the 2$^3$S,
where electrons tend to accumulate (Osterbrock~\cite{O89}),
with two important consequences.

First, photons emitted in transitions ending on 2$^3$S can
be reabsorbed, and, eventually, reemitted in different transitions. 
This self-absorption process alters the pure recombination line intensities,
increasing or decreasing them according to the line considered.

Second, collisions with free electrons may remove electrons from the 2$^3$S level 
and populate other levels, enhancing the intensities
with respect to the pure recombination value.
The most affected lines are triplets,
but singlets are also enhanced.

Self-absorption effects depend strongly on density, and 
are generally more important in planetary nebulae
than in H{\sc~ii} regions 
(Robbins~\cite{R68}; Peimbert~\cite{P95}; Peimbert, Luridiana, \& Torres-Peimbert~\cite{PLTP95}).
Collisional rates depend both on density and temperature.
Expressions for the collisional contribution to
each line can be found in Kingdon \& Ferland~\cite{KF95} and 
Benjamin {\sl et al.}~\cite{BSS99};
work on this topic has been made by, e.g.,
Ferland~\cite{F86}, Peimbert \& Torres-Peimbert~\cite{PTP87}, 
Clegg~\cite{C87}, Sawey \& Berrington~\cite{SB93}, 
Benjamin, Skillman, and Smits~\cite{BSS99}.
Both effects must be subtracted out of the total intensities
before deriving $Y$ from the observed line intensities.

\section{$Y_P$ determinations: sources of uncertainty}

\paragraph{Atomic parameters}
Benjamin {\sl et al.}~\cite{BSS99} 
identified three error sources affecting the analysis of emission lines:
a) the use of a fitting function to represent the emissivity, introducing 
an uncertainty $\sigma_{fit}$;
b) the uncertainty in the atomic data, $\sigma_{atomic}$;
c) the uncertainty in the input density and temperatures used
in the analysis, $\sigma_{n}$ and $\sigma_{T}$.
These four $\sigma$s should be added in quadrature,
and the result of these estimation should be further added
to the observational uncertainty.
These authors believe that $\sigma_{atomic}$ alone can be as high as $0.015$.

\paragraph{Underlying absorption}
The nebular diagnostics used in abundance determinations
work under the assumption that the spectrum observed
is produced exclusively in the gas.
In most cases, however, it includes also a stellar contribution;
if this superposed stellar spectrum has absorption features, the corresponding
emission lines will appear weaker than their true nebular value, introducing a bias
in the analysis if no correction is applied.

\paragraph{Ionization structure}
The total helium abundance in mass, $Y$, can be computed from the number ratio
He/H, which is, as a first approximation, equivalent to the (inferred from observations) 
ionic ratio {He$^+$}/{H$^+$} 
(or {[He$^+$+ He$^{++}$]}/{H$^+$} in high-excitation objects).
This is equivalent to assume that the Str\"omgren spheres of He and H are coincident.
When highly precise measurements are required, however,
the observed ionic ratios must be corrected
to account for either neutral helium inside the H$^+$ sphere,
or neutral hydrogen in the He$^+$ sphere.
This is generally done by multiplying the observed {He$^+$}/{H$^+$}
by an appropriate ``ionization correction factor'' ($icf$),
defined by the expression ${\rm He}/{\rm H}=icf \times {\rm He^+}/{\rm H^+}$
(alternative definitions of the ionization correction factor also exist, e.g.: 
${\rm He}/{\rm H}=[1+icf] \times{\rm He^+}/{\rm H^+}$,
or the capitalized $ICF$ by Gruenwald {\sl et al.}~\cite{GSV02}:
${\rm He}/{\rm H}=ICF \times [{\rm He}^+ + {\rm He}^{++}]/{\rm H^+}$).

\paragraph{Collisional excitation of Balmer lines}
Balmer hydrogen lines can be enhanced through collisional excitation of H$^0$.
Since this mechanism depends strongly on temperature,
it plays a role only in hot, low-metallicity regions,
where it can enhance the strongest Balmer lines by a few percent.
Davidson \& Kinman~\cite{DK85} drawed the attention to the
fact that this mechanism could introduce a bias in the measurement of $Y$:
if the Balmer flux is intepreted in terms of pure recombination,
the inferred relative abundance of hydrogen is biased toward high values,
and the He/H ratio is biased toward low values.
A further problem is that, because collisions affect more H$\alpha$ 
than H$\beta$, they mimic the effect of reddening.

\paragraph{Temperature structure}
The concept of temperature fluctuations was first introduced
by Peimbert~\cite{P67},
who developed a formalism to describe the departures from spatially constant 
temperature in nebulae, estimated their impact on abundance determinations,
and provided tools to detect their observational signatures.
Since the emissivity of each line has a unique dependence on temperature,
each line weighs differently those 
parts of the nebula with different temperatures. 
As an example, the emissivity of a collisional line such as [O{\sc~iii}] $\lambda 5007$
is, in the typical range of nebular temperatures, 
a strongly increasing function of the temperature,
therefore the observed intensity of $\lambda 5007$ 
is dominated by the hottest parts of the nebula,
and defines implicitly a typical O$^{++}$ temperature $T_e$(O{\sc~iii}).
Analogously, the recombination emissivities of  hydrogen or helium lines 
are mildly decreasing functions of the temperature,
and these lines sample preferentially the coldest zones:
their observed intensity define typical recombination
temperatures, e.g. $T_e$(He{\sc~ii}).
(Collisional contribution to these lines slightly complicates this basic picture,
because it increases with temperature; for the line as a whole, then,
the way it weighs the nebula depends on the particular regime of the object.
For example, Balmer lines in typical
nebular conditions are always dominated by recombination.) 
From the explanation above, it is clear that 
$T_e$(O{\sc~iii}) and $T_e$(He{\sc~ii}) need not take the same value,
and often indeed do not.
When we step back from intensities to abundances,
the temperature appropriate to each ion must be used to evaluate the average emissivities,
or a bias will be introduced.
This bias usually yields
spuriously low $Z$ values,
while the effect on $Y$ is more complex to predict 
since it is the combination of the opposing effects 
on the recombination and the collisional contributions.

\section{Y$_P$ determinations: results}

In this section, I will describe and compare a few recent $Y_P$ determinations.
I will use the series of works published by Izotov's group
both as a starting point, and a reference in the comparison.
This choice is motivated by two reasons. One is practical: 
their very large sample of objects has been re-analyzed by several
other groups, so the comparison is straightforward in these cases.
The other is methodological: they generally
provide an extremely detailed report of their assumptions and computations,
down to a very basic level
(with the exception, perhaps, of the uncertainties in their line intensities,
which are quoted to be extremely low and would therefore call for
an explicit discussion),
so that their results are highly reproducible.
This is a very valuable aspect of their work,
especially considering the tiny quantitative differences
among results from different authors.

Izotov and his collaborators analyzed in a series of papers 
a large sample of metal-poor BCGs
(Izotov {\sl et al.}~\cite{ITL94,ITL97},
Izotov \& Thuan~\cite{IT98}, hereinafter ITL94,97 and IT98).
These works discuss critically 
the potential bias introduced by several physical effects,
of which a few will be mentioned here.
The amount of {\bf stellar absorption} is determined by ITL94
simultaneously with the reddening coefficient, by fitting iteratively
the dereddened intensity of several hydrogen lines to their recombination values.
They find that stellar absorption is generally negligible;
however, their procedure fails, for no evident reason,
for two of the regions.
On the contrary, stellar absorption is found by ITL97 to be extremely
important in the case of {I Zw 18} (a crucial object in $Y_P$ determinations
since it is the most metal-poor galaxy known),
which is therefore excluded from the analysis.
As for the {\bf ionization structure}, ITL94 compute the
$icf$s by means
of a simple recipe by Pagel {\sl et al.}~\cite{Pal92a}, 
linking the $icf$ to $\eta_{soft}$\footnote{I added the subindex {\it soft} to the customary symbol to avoid confusion with the cosmological $\eta$.}
(V\'\i lchez \& Pagel~\cite{VP88}),
and  corroborate the result with a fit to the photoionization models by Stasi\'nska~\cite{S90}.
They find $icf$s$ \greatsim 1$ for the objects in the sample,
but in IT98 the question is re-analyzed and the $icf$s are set to 1.
As for the {\bf temperature structure}, 
ITL94 adopt $T_e$(He{\sc~ii})=$T_e$(O{\sc~iii}), 
based on a fit to the models by Stasi\'nska~\cite{S90}.
ITL97 maintain this assumption and exclude,
based on several indirect pieces of evidence, 
that temperature fluctuations might play a role in the objects considered.
On the other hand, ITL94
claim that a proper estimation of collisions in helium lines 
should rely on a self-consistent density value, $N_e$(He{\sc~ii}), rather than the arbitrary assumption
of a density obtained by other diagnostics.
They calculate $N_e$(He{\sc~ii}) by means of a self-consistent procedure,
which constrains the three He{\sc~i} line ratios $\lambda$5876/$\lambda$4471, 
$\lambda$6678/$\lambda$4471, and $\lambda$7065/$\lambda$4471
to recover their recombination value after correcting for collisional enhancement.
The importance of including the density-sensitive $\lambda$7065 is stressed
as a means to obtain a self-consistent result.
{\bf Self-absorption} effects are considered by ITL94 to be negligible, on the
argument that the most sensitive line, He{\sc~i} $\lambda$3889,
has roughly its recombination value.
However, the analysis performed by IT98 on a larger sample leads
to the conclusion that self-absorption effects are indeed important,
and $\lambda$3889 is explicitly added to their self-consistent procedure
to detect self absorption.
{\bf Collisional effects on hydrogen lines} are evaluated by ITL97, but
because the inclusion of such effects actually worsens the fit, 
the authors infer that they are probably overestimated, and choose not
to include them in the analysis.
The last paper of this series analyzes a sample of 45 BCGs,
yielding a primordial helium value of $Y_P=0.244\pm 0.002$.

Let's see now how these effects have been treated
by other authors.
Several of them have centered their analysis on the
{\bf ionization correction factor}:
{\bf a)} Olive \& Steigman~\cite{OS95} agree with ITL94 in that $icf\sim 1$.
{\bf b)} Based on photoionization models, Viegas {\sl et al.}~\cite{VGS00} 
argue that $icf\lesssim 1$ in regions ionized by young, metal-poor stars, 
so that helium abundances derived in previous analyses should
be corrected downwards; this effect is amplified
by density inhomogeneities. 
Re-analyzing the sample of IT98, they find $Y_P=0.241\pm 0.002$. 
{\bf c)} Ballantyne {\sl et al.}~\cite{BFM00} find, 
by means of photoionization models, 
that at high stellar temperatures
the $icf$s can be significantly different from 1,
with both negative or positive values possible according to the
particular stellar atmosphere and temperature considered.
They propose to use the metallicity-independent line ratio
$\lambda\,5007$/$\lambda\,6300$
to discriminate the regions for which $icf\ne 1$.
Applying this exclusion criterion to the IT98 sample, 
they find $Y_P=0.2489\pm0.0030$.
{\bf d)} Sauer \& Jedamzik~\cite{SJ02} consider 
the ionization structure a major source
of uncertainty in the determination of $Y_P$, and,
by means of photoionization models,  develop a method to determine
the $icf$ based on $\eta_{soft}$. 
They find characteristic $icf$s values smaller than 1
for the sample by IT98 and, 
though they don't give any definite numbers,
conclude that $Y_P$ was overestimated by those authors.
{\bf e)} Gruenwald {\sl et al.}~\cite{GSV02} investigate the evolution 
of the $icf$ as the H{\sc~ii} region evolves, 
and find that in the range of ages in which H{\sc~ii} regions
are observed, the $icf$ oscillates twice back and forth
from negative to positive values.
They argue that the criterion proposed 
by Ballantyne {\sl et al.}~\cite{BFM00} is not sensitive
to the shape of the ionizing spectrum, but rather to its intensity.
They also argue that partially density-bounded regions may
have high $\lambda\,5007$/$\lambda\,6300$ ratios, mimicking 
a high-excitation zone and biasing the application of the criterion.
Re-analyzing the data by Izotov \& Thuan~\cite{IT98}, they find that 
$Y_P$ should be lowered to $Y_P=0.238\pm0.003$.

Other authors have focused their attention on the treatment 
of {\bf self-absorption}:
for example, 
Olive, Steigman \& Skillman~\cite{OSS97}
argue against the use of $\lambda$7065 by ITL94 and ITL97.
These authors believe that, because
$\lambda$7065 is very sensitive to self-absorption effects, for which no correction has been done,
 it may introduce 
large uncertainties in the results.

The treatment of {\bf underlying stellar absorption} has been
carried out differently by different authors.
{\bf a)} Olive \& Skillman~\cite{OS01} believe that this effect 
might play a role,
and propose to include He{\sc~i} $\lambda 4026$ in the self-consistent 
analysis of helium lines: this line could serve 
as a diagnostic of underlying absorption, since it is
weak and not much affected
by collisions and self-absorption.
{\bf b)} Peimbert {\sl et al.}~\cite{PPL02} correct
the weakest helium lines 
for underlying absorption according to the 
synthetic spectra by Gonz{\' a}lez Delgado, Leitherer, \& Heckman~\cite{GDLH99}.

A few authors have discussed the
question of the {\bf temperature structure}.
{\bf a)} Steigman {\sl et al.}~\cite{SVG97} argue that temperature
fluctuations bias differently 
hot, low-$Z$ than cold, high-$Z$ regions.
The net effect of taking temperature
fluctuations into account would be to tilt the $Y$ vs. $O$ relation,
in the sense of making it flatter. 
{\bf b)} The temperature structure is the main theme in the 
work of A. Peimbert, M. Peimbert and collaborators~\cite{PPR00,PPL02}.
These authors argue, based on several lines of evidence
from observations and photoionization modeling, 
that in low-metallicity regions $T$(He{\sc~ii}) 
is systematically smaller than $T$(O{\sc~iii}).
They analyze a small sample of metal-poor objects,
and determine $T_e$(He{\sc~ii}) and $N_e$(He{\sc~ii})
self-consistently by means of a $\chi^2$ minimization procedure
applied to the intensity of up to nine helium lines,
obtaining on average {$T_e$(O{\sc~iii}) -- $T_e$(He{\sc~ii}) = 1300 K.}
The primordial helium abundance they determine is $Y_P=0.2384\pm0.0025$.

Peimbert {\sl et al}~\cite{PPL02} also evaluate
the {\bf collisional enhancement of Balmer lines},
which acts in the sense of increasing their
computed value of $Y_P$ by about 0.003 (this increase is
already included in the $Y_P$ value quoted above).
This effect has also been studied by
Stasi{\' n}ska \& Izotov~\cite{SI01}, who 
estimate that the correction for individual
objects can be as high as 5\%, making it
one of the most important sources of systematic errors
in the determination of $Y_P$. 

\section{Conclusions}

From the discussion above, it is apparent that
the central problem in the determination of $Y_P$ is 
that several physical mechanisms acting in H{\sc~ii} regions 
are still not completely understood.
Furthermore, although I described them separately for exposing convenience,
these mechanisms interact with each other in complex ways.
A huge collective effort is presently aiming 
to pinpoint the relevance of these effects,
in part with direct observations, more often
with numerical simulations.
In the meanwhile,
whether they actually play a role or not 
remains mostly a question of personal judgement, based on
pieces of evidence that are more or less compelling, but 
rarely conclusive.
Because personal judgement is so important,
it is a natural question whether
unconscious individual prejudices might be playing a role in obtaining one result or another.
It is well known that, to some extent, this kind of bias 
is always present in any analysis, and that it may be
particularly insidious. I therefore conclude with a 
personal remark: it would be extremely instructive for all of us
if the relevant scientists in the field 
build up 
a kind of double-bind experiment, with both real and ``placebo'' data,
to evaluate the impact of the human factor on $Y_P$ determinations. 

\section*{Acknowledgments}
This research has been supported by a Marie Curie Fellowship
of the European Community programme {\sl ``Improving Human Research Potential and the Socio-economic Knowledge Base''} under contract number HPMF-CT-2000-00949.

\end{document}